\documentclass[aps,showkeys,pre,reprint,superscriptaddress,groupaddress,floatfix,nofootinbib]{revtex4-2}
\usepackage{graphicx}
\usepackage{soul}
\usepackage{dcolumn}
\usepackage{bm}
\usepackage[utf8x]{inputenc}
\usepackage{amsmath}
\usepackage{amsfonts}
\usepackage{amssymb}
\usepackage{graphicx}
\usepackage[dvipsnames]{xcolor}
\usepackage[T1]{fontenc}
\usepackage[normalem]{ulem}

\usepackage{braket}

\begin{document}
\title{Complex first-passage transport in ring networks with long-range jumps and stochastic resetting}

\author{Oscar Ivan \surname{Torres Mena}}
\email{oscart@estudiantes.fisica.unam.mx}
\affiliation{Instituto de F\'isica, Universidad Nacional Aut\'onoma de M\'exico, Apdo.\ Postal 20-364, 01000, Ciudad de M\'exico, México}

\author{Francisco J. Sevilla}
\email{fjsevilla@fisica.unam.mx}
\thanks{corresponding author}
\affiliation{Instituto de F\'isica, Universidad Nacional Aut\'onoma de M\'exico, Apdo.\ Postal 20-364, 01000, Ciudad de M\'exico, México}

\begin{abstract}
The transport properties of discrete-time random walks on ring networks with deterministic shortcuts are investigated through analytical and numerical methods. The network consists of a periodic chain where each node is connected to its nearest neighbors and to nodes located at a fixed distance 
$r$. Using the spectral properties of the transition matrix, we derive explicit expressions for the occupation probabilities and mean first-passage times (MFPTs). Contrary to the common expectation that shortcuts monotonically enhance transport, we find that the MFPT between distant nodes develops a highly non-monotonic dependence on the shortcut length. Beyond a threshold value, the MFPT landscape exhibits a hierarchy of maxima and minima organized in a self-similar pattern associated with commensurability relations between the shortcut length and the system size. The scaling behavior of these extrema reveals regimes where transport efficiency is either strongly enhanced or suppressed. We further analyze the mean squared displacement and the influence of stochastic resetting, showing that resetting amplifies the oscillatory MFPT structure and induces strongly nonuniform stationary distributions. These results demonstrate that the spatial organization of long-range connections plays a crucial role in determining transport efficiency in networks.
\end{abstract}
\maketitle

\section{Introduction}
The study of discrete-time random walks (DTRW) on networks, either regular or complex, has established a paradigmatic basis to analyze the dynamics of a great variety of process. Fundamentally, the \emph{random walk} models the stochastic journey of a system (the walker), over its different accessible states (the network nodes). This stochastic dynamics requires the knowledge of the transition probabilities between connected pairs of the system states, the set of connected pairs defines the set of edges or links between the set of the system states (nodes). These three elements define the dynamics of walks on networks \cite{lovasz1993random, hughes1995random, chung1997spectral,levin2017markov,noh2004random,barrat2008dynamical,rosvall2008maps,burda2009localization,newman2010networks,sinatra2011maximal,lambiotte2011flow,masuda2017random}. Of relevance are the transport properties of free walks or walks under \emph{stochastic resetting}, and questions as: How long does it take to the system to go from one node to another one? How fast is the transport process? Or how fast the dispersion from the average grows with time? have been fundamental for the understanding of the system dynamics. Reciprocally, networks can be characterized by the transport properties of random walks on them \cite{polizzi2016mean, zaman2023kemeny, Eraso2024Antifragility}. 

The \emph{first passage time} (FPT) of a stochastic process has been a relevant quantity, since this marks the waiting time for which the system reaches a state of interest for the first time. In the context of random walks on networks, the FPT problem refers to the calculation of the probability distribution of the time it takes to the walker to reach a given node when it starts its walk from another one. This problem has been widely analyzed and relevant results are known, particularly in the simplest case of a one-dimensional random walk on a ring network of sites connected by nearest neighbors (linear chain with periodic boundary conditions) \cite{Montroll1964, montroll1965randomII,WeissJStatPhys1981,masuda2017random}. One relevant result is that the mean first-passage time (MFPT) between a pair of nodes, scales with the ring-network size, $N$, as $N^2$. Thus, MFPT diverges quadratically with $N$ as $N \rightarrow  \infty$.
Of recent interest have been a series of different dynamics that modifies the distribution of FPTs for a ring-network random walk, particularly when the ring network is modified into the model of Newman and Watts that give rise for the small-world properties \cite{watts1998collective, parris2005traversal}. The global mean-first passage properties of a continuous-time random walk to a target site, averaged over its starting position, were analyzed in Ref. \cite{parris2005traversal, parris2008random, candia2007transport} when the transition rate between first neighbors is distinct from steps across long connections. There, it is shown that MFPT can be reduced by modulating the number of shortcuts and the transition rates between them, and a lower bound for the global mean-first passage time is found. These results have been extended to general complex networks in Ref. \cite{Tejedor_2009}.

Transport processes on networks with long-range connections have attracted considerable attention due to their relevance in physical and technological systems. The presence of shortcuts is commonly associated with the small-world effect, where the addition of a small number of long-range links drastically reduces typical path lengths and enhances transport efficiency. In many models, this intuition leads to the expectation that increasing the range of long-distance connections should monotonically accelerate diffusive exploration. However, the interplay between lattice periodicity and long-range transitions may lead to more subtle behavior. In particular, when long-range connections introduce additional spatial scales into the dynamics, resonant or commensurate structures may emerge that modify transport properties in nontrivial ways. Understanding how such mechanisms influence first-passage processes remains an open question even in relatively simple network topologies.

\begin{figure}
\includegraphics[width=0.33\columnwidth]{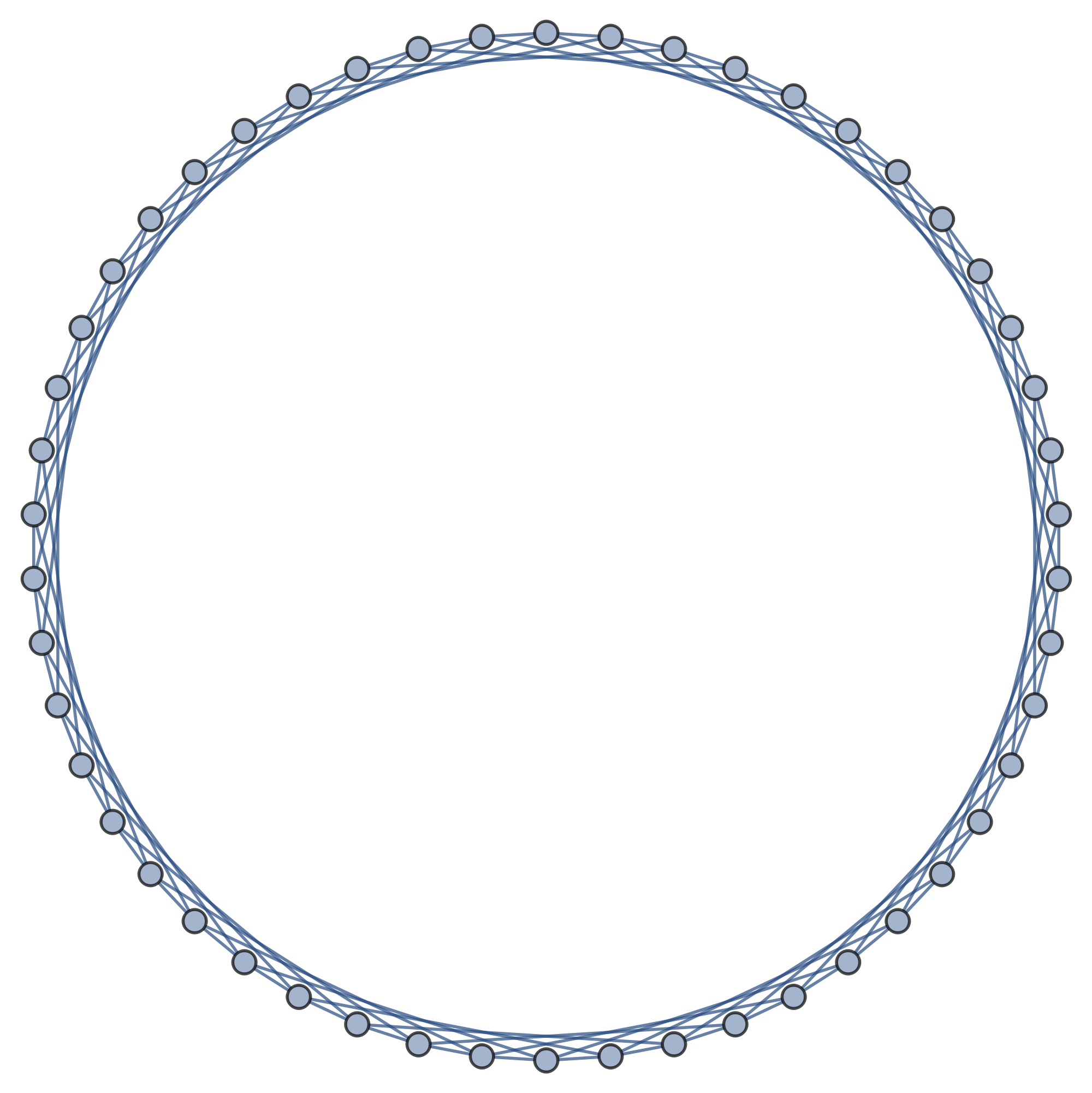}\includegraphics[width=0.33\columnwidth]{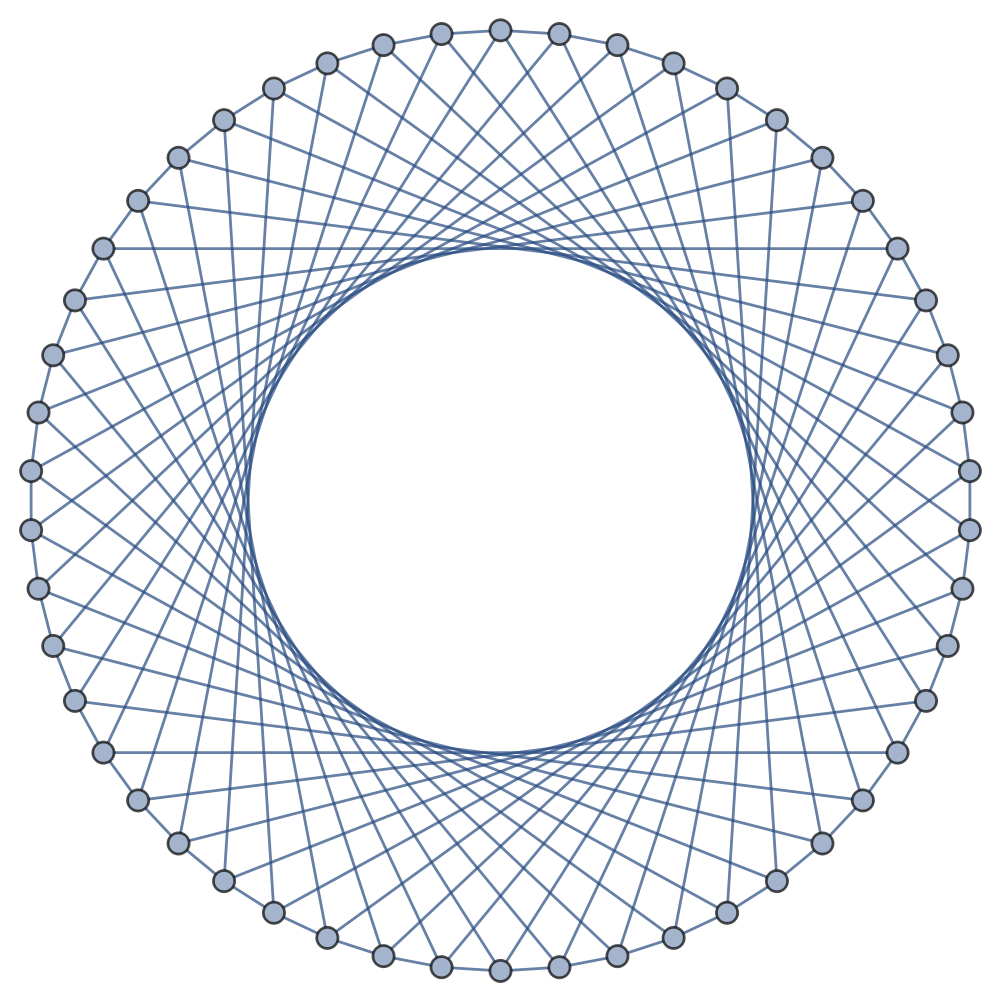}\includegraphics[width=0.33\columnwidth]{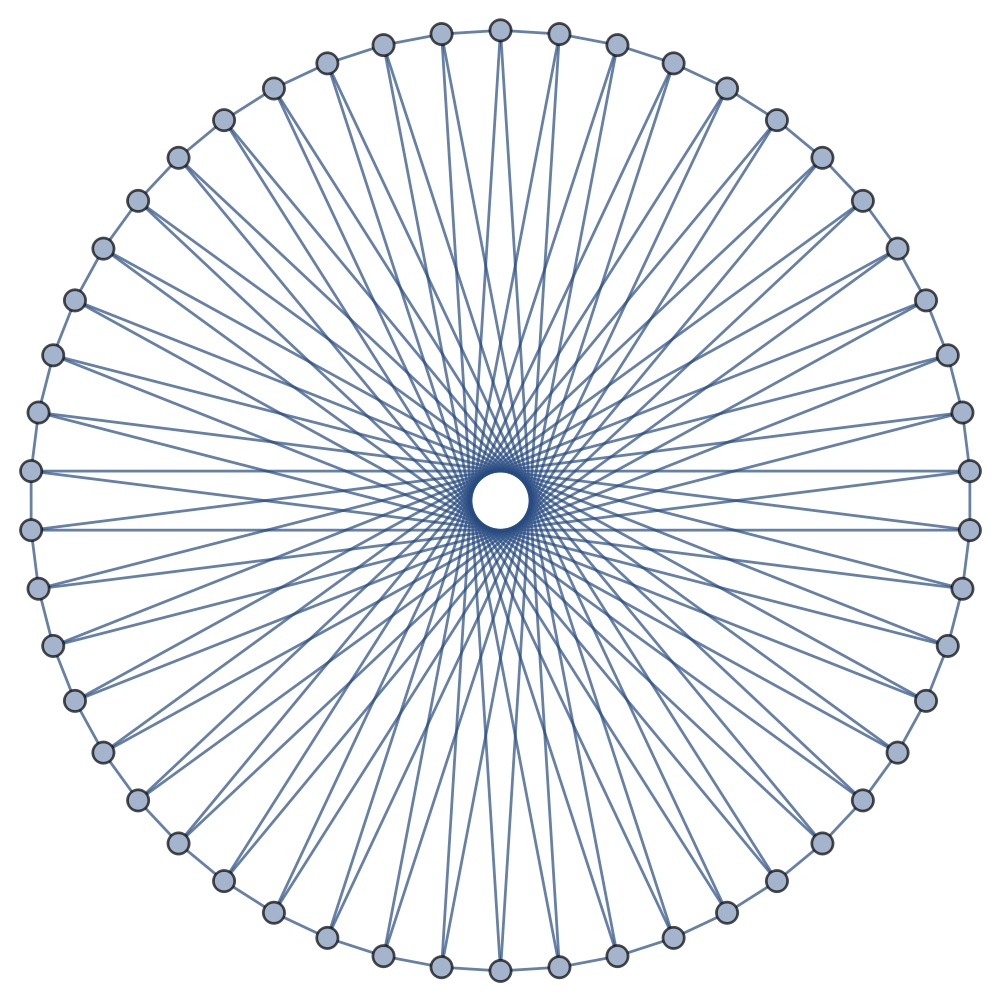}
\caption{Three examples of the networks used in this paper, each node $j$ is connected to its two nearest neighbors $j\pm1$ and to the two nodes located at a distance $r$, $j\pm r$, with $1<r<N/2-1$. The ring size is $N=50$ and $r$ (left to right) takes the values $5$, $16$ and $24$.}
\label{fig:Figure1}
\end{figure}

In this work we investigate this question using a minimal but analytically tractable model: a ring network in which each node is connected to its nearest neighbors and to nodes located at a fixed distance $r$ (see Figure \ref{fig:Figure1}). This system allows us to explore how the interplay between local diffusion and deterministic long-range jumps affects transport properties. 

A central aspect of the results reported here is the emergence of a highly non–monotonic dependence of the mean first-passage time on the shortcut length. This behavior can be understood as a consequence of the spectral structure of the transition matrix governing the dynamics. The eigenvalues of the stochastic matrix contain two periodic contributions associated with local hops and long-range shortcuts, which generate a competition between two characteristic spatial frequencies. When the shortcut length becomes commensurate with the distance between the nodes involved in the first-passage event, constructive or destructive interference between these spectral modes occurs. As a consequence, transport may be either strongly enhanced or suppressed depending on the value of the shortcut length. This mechanism explains the appearance of the alternating sequence of maxima and minima in the MFPT landscape and provides a physical interpretation of the self-similar structures observed in the numerical results. 
In addition, we consider the influence of stochastic resetting on the random walk and show that the stationary configurations induced by the resetting clearly reveals the network structure, which have recently elucidated different characteristics of networks \cite{BonomoPhysRevE2021,ChenPhysRevE2022}. 

 Although the analysis presented here focuses on a deterministic shortcut ring, the mechanism responsible for the observed MFPT structure does not rely on the specific details of this topology. The key ingredient is the coexistence of two spatial scales in the transition dynamics: local hops and long-range jumps. Similar situations arise in networks with heterogeneous long-range connections, Lévy-like random walks, and small-world architectures. Therefore, the ring model studied here should be regarded as a minimal setting that isolates the fundamental mechanism responsible for the transport anomalies observed.

The present paper is organized as follows: in section \ref{sec:MFPT} the theoretical framework is briefly presented, in section \ref{sec:transport_propierties} are presented results related to transport properties without stochastic resetting, whose effects are explored in section \ref{sec:stochastic_resetting}, and finally in section \ref{sec:conclusions} the conclusions.

\section{The system under study and theoretical framework} \label{sec:MFPT}

We consider a discrete-time random walk on an $r$-shortcut ring with $N$ nodes, which corresponds to a linear chain of nodes $i=1,\,\ldots,N$, with periodic boundary conditions, i.e., $ 
 i\pm N\rightarrow i$.  This regular network of degree $k=4$, is built by connecting each node to its two nearest neighbors by unit-length links (simply called links) and to two nodes at distance $r$ by links of length $r$ (shortcuts). Additionally we assume that the DTRW dynamics is  homogeneous, i.e., we consider that the transition probability of the walker from one site to its four neighbors is constant and the same (undirected network), this is enough to exhibit a complex dynamics, however some generalizations are straightforward. We will show in this paper that the transport properties of a random walk in these networks are not trivial and strongly depend on the shortcut length $r$.

The probability $P_{s, j}(t)$, of finding the walker at site $j$ after $t$ steps, given that the walk started at site $s$ is given by the solution of the master equation
\begin{equation}
    P_{s, j}(t)=\sum_{m} P_{s, m}(t-1) \pi_{m, j}, 
    \label{eq:master}
\end{equation}
where $\pi_{i,j}$ denotes the probability of transiting from node $i$ to node $j$ in one step, and are the entries of a \emph{stochastic matrix}, also called the \emph{transition matrix}, $\Pi$. These satisfy, by definition, that $\sum_j\pi_{i,j}=1$. The eigenvalues and eigenvectors of $\Pi$ allow for the calculation of the occupation probabilities $P_{s,j}(t)$, including the
stationary occupation $P_j^\infty$, as well as the MFPT to any node. 

If transitions along links occur with probability $p$, and transitions along shortcuts with probability $q$, 
the entries of $\Pi$ are given by $\pi_{i,j}=p[\delta_{i,j-1}+\delta_{i,j+1}]+q[\delta_{i,j-r}+\delta_{i,j+r}]$ and therefore $2p+2q=1$.
It is clear therefore that
\begin{equation}
\label{TransitionMatrix}
    \Pi = p \Pi_l + q \Pi_s,
\end{equation} 
where the stochastic matrices $\Pi_l$ and $\Pi_s$ describe the transitions along links and shortcuts, and have entries $[\Pi_l]_{i,j}=\delta_{i,j-1}+\delta_{i,j+1}$ and $[\Pi_s]_{i,j}=\delta_{i,j-r}+\delta_{i,j+r}$, respectively. Due to the ring topology, both matrices $\Pi_l$ and $\Pi_s$ are circulant (matrix rows are defined by the cyclic right shift of the first row), however, in general $\Pi$ is not circulant in itself. Some transport properties of random walks on ring networks have been studied using this property \cite{Riascos_2015,Allen-Perkins_2019,tejedor2010response}. 

The eigenvalues $\{\lambda_{\mu}\}$ and eigenvectors $\ket{\Phi_\mu}$ of a $N\times N$ circulant matrix are known to be
\begin{subequations}
\label{EigenSystem}
 \begin{equation}
    \lambda_{\mu}=\sum_{k=0}^{N-1} a_{k+1} \exp \left( {\mathrm{i}\frac{2 \pi}{N}\mu\,k} \right),\, \mu=0,\ldots,N-1,
\end{equation}
with $a_{k}$, $k=1,\ldots,N$, are the elements of the first row of the circulant matrix; while the corresponding normalized eigenvectors are  
\begin{equation}
    \ket{\Phi_\mu}=\frac{1}{\sqrt{N}}\left(
    \begin{array}{ll}
        1\\
		\exp \left({\textrm{i}\frac{2 \pi}{N}\mu} \right)\\
        \vdots\\
        \exp \left( {\textrm{i}\frac{2 \pi}{N}(N-1)\mu} \right)
    \end{array}
    \right),\, \mu=0,\ldots,N-1,
\end{equation}
which do not depend on the $a_k$'s and therefore are common to all circulant matrices \cite{Allen-Perkins_2019}. This is a well-known result in condensed matter physics, where the diagonalization of translation-invariant operators, which are circulant matrices in the representation of discrete space, is done via Fourier modes \cite{kittel1963quantum}. This property will allow us to compute the eigenvalues of the complete transition matrix $\Pi$, since 
\begin{align}
\Pi\ket{\Phi_{\mu}}=&p\Pi_{l}\ket{\Phi_{\mu}}+q\Pi_{s}\ket{\Phi_{\mu}}\nonumber\\
=&(p\lambda_{l,\mu}+q\lambda_{s,\mu})\ket{\Phi_{\mu}},
\end{align}
\end{subequations}
where $\lambda_{l,\mu}$ and $\lambda_{s,\mu}$ are given by
\begin{subequations}
\begin{align}
\lambda_{l,\mu} &= \left( \exp \left({\textrm{i}\frac{2 \pi }{N}\mu}\right) + \exp \left( {\textrm{i}\frac{2 \pi}{N}\mu(N-1)}\right) \right)\nonumber\\
    &=2 \cos\left(\frac{2 \pi \mu}{N}\right),\\
\lambda_{s,\mu} &= \left(\exp \left({\textrm{i}\frac{2 \pi }{N}\mu\, r} \right) + \exp \left({\textrm{i}\frac{2 \pi }{N}\mu(N-r)} \right) \right)\nonumber\\
&=2\cos\left(\frac{2 \pi \mu r}{N}\right),
\end{align}
\end{subequations}
respectively. Thus, the eigenvalues of the complete transition matrix $\Pi$ are real and given by
\begin{equation}
    \lambda_{\mu}=2p \cos\left(\frac{2 \pi \mu}{N}\right) + 2q \cos\left(\frac{2 \pi \mu r}{N}\right),
    \label{eq:eigenvalues}
\end{equation}
which satisfy $|\lambda_\mu|\le1$ for $\mu=0,\ldots,N-1$. 

The use of Dirac's notation simplifies the presentation of some calculations. Following \cite{Riascos_2020} we have that equation \eqref{eq:master} can be written as
\begin{equation}
P_{s,j}(t)=\bra{s}\Pi^t\ket{j},
\end{equation}
where the set of $N$ vectors $\{\ket{j}\}$ denotes the $N$-dimensional canonical basis and $\bra{j}$ its corresponding transpose. 
Since in our particular case $\Pi$ is real and symmetric it has the diagonal representation $\sum_{\mu=0}^{N-1}\lambda_{\mu}^{t}\ket{\Phi_\mu} \bra{\Phi_{\mu}}$, thus  
\begin{equation}
    P_{s,j}(t)=\sum_{\mu=0}^{N-1}\lambda_{\mu}^{t}\braket{s|\Phi_{\mu}} \braket{\Phi_{\mu}|j},
    \label{eq:probabilidad_s_j}
\end{equation}
where the term $\braket{s|\Phi_\mu}$ is the projection of the state $\ket{s}$ onto the eigenvector $\ket{\Phi_{\mu}}$, similarly, $\braket{\Phi_\mu|j}$is the projection of $\ket{\Phi_{\mu}}$ onto the basis state $\ket{j}$. 

The stationary distribution $P_j^\infty=\lim_{t\rightarrow\infty}P_{s,j}(t)$, when it exists, is relevant in the calculation of different transport properties, in particular for the MFPT. The existence of such a limit is guarantied under the  irreducibility and a periodicity of the transition matrix $\Pi$ \cite{LevinBook2017}. In such a case, $P_j^\infty$ is given in terms of the normalized eigenvector $\ket{\Phi_0}$ for which $\lambda_0=1$ according to Eqs.~\eqref{EigenSystem}. In the numerical cases considered here, we have chosen rings with an even number of nodes, thus for odd $r$ the $N/2$-th eigenvalue is $\lambda_{N/2}=-1$, which  indicates that rings with an even number of nodes and shortcuts of length odd, the random walk is of period 2. Additionally, the inverse of the spectral gap $g=1-\lambda_\star$, also called the \emph{relaxation time} $\tau_\star$, measures how fast the probability distribution converges to the stationary state \cite{masuda2017random,LevinBook2017}, where $\lambda_\star$ is the absolute value of the second largest eigenvalue. For large $N$ and $r\ll N$, $\lambda_\star=|\lambda_1|\approx1$ and the convergence to stationary distribution is rather slow, as $\tau_\star$ scales as
$\tau_\star\sim\frac{N^2}{4}\frac{1}{\pi^2 (p+qr^2)}$. For different values of $r$ we have that $\lambda_\star\neq|\lambda_1|$, but still $\lambda_\star\approx1$ thus the scaling $\tau_\star\sim\frac{N^2}{4}$ remains.

 \section{Transport properties}
 \label{sec:transport_propierties}
We now present the analysis of the transport properties of a Markovian random walk on the network considered here. We focus mainly on the dependence of the MFPT between pairs of nodes, say $s$ and $j$, 
$\braket{\tau_{s,j}}$, on the shortcuts length $r$. We also analyze the evolution of the mean squared displacement (MSD). 

\subsection{First-passage time statistics}
\label{sec:results_mfpt}

By resorting to the following identity 
\begin{equation}
    P_{s,j}(t)=\delta_{s,j}\delta_{t,0}+\sum_{t'=0}^{t} P_{j,j}(t-t')F_{s,j}(t),
    \label{eq:FPT_dis}
\end{equation}
that involves the probability density, $F_{s,j}(t)$, of the FPT of arriving at node $j$ departing from node $s$. The first term in the last equation gives the probability of finding the walker at the initial node $s$, while $P_{j,j}(t)$ is the probability density of returning to node $j$ in $t$ steps when departing from itself. The MFPT time is 
\begin{equation}
\label{eq:Figure2}
\braket{\tau_{s,j}}=\sum_{t=1}^{\infty} t F_{s,j}(t).
\end{equation}
The standard method of analysis is the \emph{generating function method}, from which the discrete Laplace transform $\bigl[\widetilde{f}(\epsilon)=\sum_{t=0}^\infty \exp({-\epsilon t})f(t)$, $\epsilon\ge0\bigr]$ is suitable for solving Eq. \eqref{eq:FPT_dis} \cite{MetzlerBook2014,masuda2017random}. The results are well known and we just briefly point out the calculation. Thus after taking the discrete Laplace transform to Eq. \eqref{eq:FPT_dis}, we have $\widetilde{P}_{s,j}(\epsilon)=\delta_{s,j}+\widetilde{P}_{s,j}(\epsilon)\widetilde{F}_{s,j}(\epsilon)$, from which we obtain
\begin{equation}
\widetilde{F}_{s,j}(\epsilon)=\frac{\widetilde{P}_{s,j}(\epsilon)-\delta_{s,j}}{\widetilde{P}_{j,j}(\epsilon)}.
\end{equation}
$\widetilde{P}_{s,j}(\epsilon)$ can be computed from \eqref{eq:probabilidad_s_j} and is given by
\begin{equation}
\label{eq:probabilidad_s_j-Laplace}
\widetilde{P}_{s,j}(\epsilon)=\sum_{\mu=0}^{N-1}\frac{\braket{s|\Phi_\mu} \braket{\Phi_\mu|j}}{1-\exp({-\epsilon})\lambda_\mu},
\end{equation}
which leads to \cite{masuda2017random, Riascos_2020, riascos2021random}
\begin{equation}
    \braket{\tau_{s, j}}=\frac{1}{P_{j}^{\infty}}\left[\delta_{s, j}+\sum_{\mu=1}^{N-1} \frac{1-\exp{(\textrm{i}\frac{2\pi (s-j)}{N}\mu)}}{1-\lambda_{\mu}} \right]
    ,    \label{eq:tau}
\end{equation}
where we have used that $\braket{\tau_{s,j}}=-\bigl[\frac{d}{d\epsilon}\widetilde{F}_{s,j}(\epsilon)\bigr]_{\epsilon=0}$ and assumed that $\lambda_0=1$ is the only eigenvalue with magnitude $1$ (this can be assured even for the cases where the random walk is not aperiodic by implementing a mathematical artifact, namely, by simply redefining the transition matrix as $\Pi^\prime=(1-\eta)\Pi+\eta\mathbb{I}$ with $\eta\ll1$ and $\mathbb{I}$ the $N\times N$ identity matrix). 

It can be shown that for networks that are invariant under shifting sites (nodes), $\braket{\tau_{s,j}}$ depends only on the the separation between the nodes involved $d=|s-j|$, i.e., $\braket{\tau_{s-j}}=\braket{\tau_{d}}$ with
$d=1,2,\ldots,N/2$.
\begin{figure}
\includegraphics[width=\columnwidth]{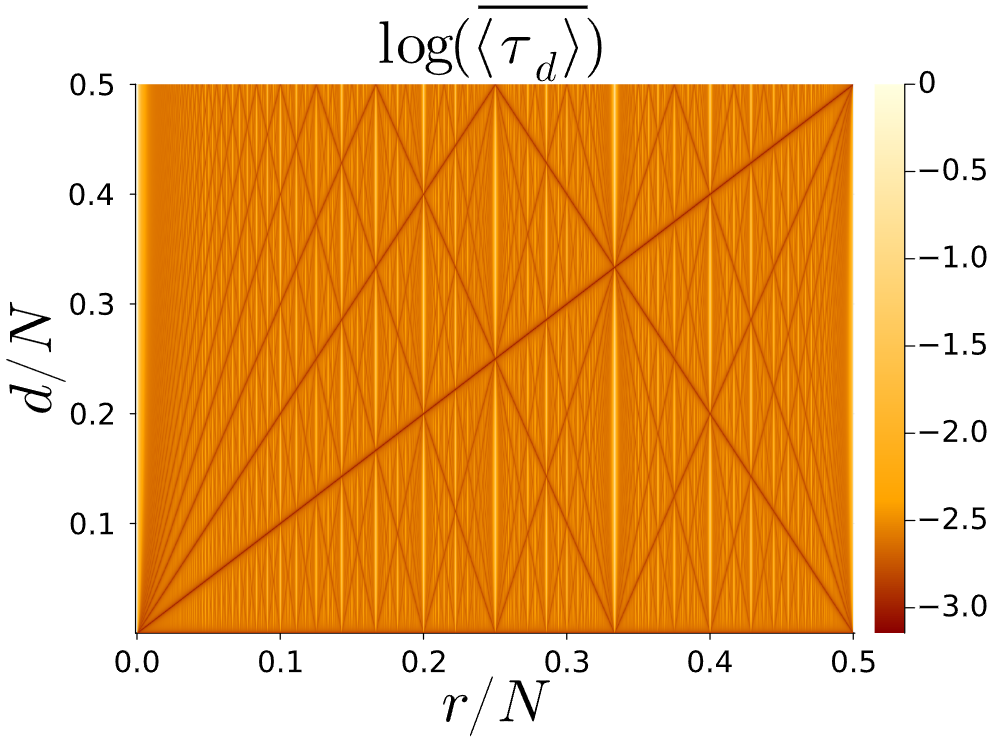}\\
\includegraphics[width=0.98\columnwidth]{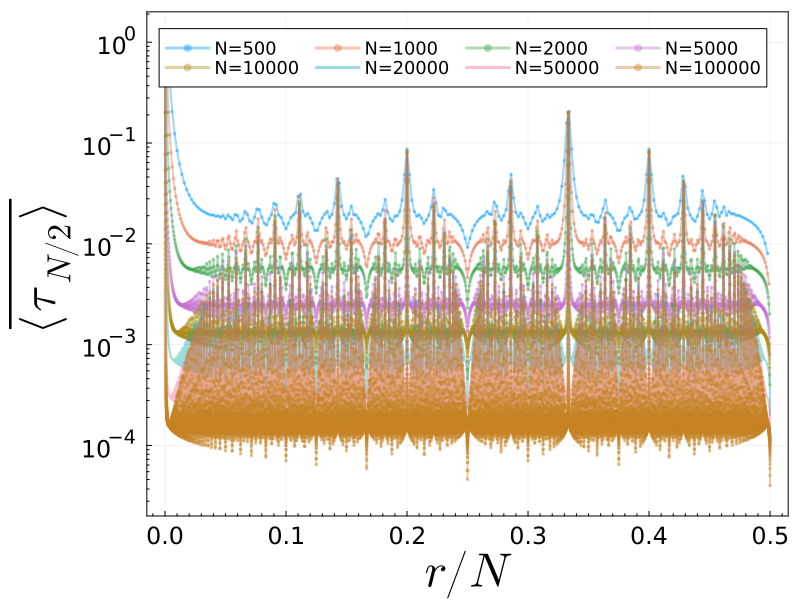}
\hspace{0.1cm}
\caption{Top panel.- Logarithm of MFPT,  $\log(\overline{\braket{\tau_{d}}})$, is shown as function of $r/N\in[1, 1/2-1/N]$ and distance  $d=|s-j|\in [1,N/2]$ 
for a ring size of $N=5000$ nodes.  Bottom panel.- The normalized MFPT $\overline{\braket{\tau_{d}}}=\braket{\tau_{d}}/(N/4)$ between nodes at distance $d=N/2$ (diametrically opposed in the ring) as function of the shortcut length $r$ for a ring sizes $N=500$, 1000, 2000, 5000, 10000, 50000 and 10000.}
    \label{fig:Figure2}
\end{figure}
From Eq.~\eqref{eq:probabilidad_s_j-Laplace} $P_j^\infty$ can be obtained by noticing that in the limit $\epsilon\rightarrow0$ the only term that contributes to the pole is the one corresponding to $\mu=0$, thus we have
\begin{equation}
P_j^\infty=\lim_{\epsilon\rightarrow0}\epsilon \widetilde{P}(\epsilon)_{s,j}=\braket{s|\Phi_{0}}\braket{\Phi_{0}|j}=\frac{1}{N},
\label{eq:pj_srs}
\end{equation}
stating the ergodic nature of the random walk dynamics considered.

In the top panel  of FIG.~\ref{fig:Figure2} the logarithm of the MFPT $\braket{\tau_{d}}$ (indicated by the color bar) as function of $d$ and of the shortcut length $r=2,3\ldots,N/2-1$ is shown (unless stated otherwise, we do not write the dependence on 
$r$ explicitly). The MFPT is normalized with $\braket{\tau_{N/2}}_{max}=N^2/4$, i.e. with the MFPT of pairs of farthest nodes; $d$ and $r$ are normalized by the system size $N$. As different results are presented by using this normalization we introduce the notation $\overline{\braket{\tau_{s,j}}}$ to indicate $\braket{\tau_{s,j}}/\braket{\tau_{N/2}}_{max}$. In the top panel the color bar indicates the values of $\log\bigl(\overline{\braket{\tau_{d}}}\bigr)$, being darker the smaller the MFPT for given $d/N$ and $r/N$. We find that the addition of shortcuts gives a rich dependence on $d$ and $r$ as is discussed afterwards. The top panel of FIG.~\ref{fig:Figure2} also exhibits clear patterns for simple correlations between $d$ and $r$ for the MFPT. The smaller MFPTs are obtained for separations $d$ between pair of nodes that coincides with the length of the shortcut, i.e. for $d=r$, in general the smaller MFPTs are observed for the cases for which the linear relations
\begin{equation}
d=\begin{cases}
  \nu\, r & \textrm{if } r\in \bigcup\limits_{i=1, \textrm{odd}}^\nu \Bigl[\frac{i-1}{\nu}\bigl(\frac{N}{2}-1\bigr),\frac{i}{\nu}\bigl(\frac{N}{2}-1\bigr)\Bigr],\\
  N-\nu\, r & \textrm{if } r\in \bigcup\limits_{i=2, \textrm{even}}^\nu \Bigl[\frac{i-1}{\nu}\bigl(\frac{N}{2}-1\bigr),\frac{i}{\nu}\bigl(\frac{N}{2}-1\bigr)\Bigr],
  \label{eq:d_minimas}
\end{cases}
\end{equation}
are fulfilled, where $\nu=1,\ldots, \frac{N}{2}-1$ denotes the commensurability index between $d$ and $r$. The finite union of the intervals in Eq.~\eqref{eq:d_minimas} contain the values of $r$ for which the relations $\nu r$ or $N-\nu r$ are valid. These cases are clearly shown as  the darken-shadowed lines in the referred figure.  
The brighter regions correspond to larger values of the MFPTs, which are of the order of the maximum value $\langle\tau_{N/2}\rangle_{max}=N^{2}/4$.

For a given value $d$, the MFPT decreases monotonically with $r$, as expected, since adding shortcuts enhances ring traversal and thus improves FPT. This regime is clearly shown for small values of $r/N$ in the bottom panel of FIG.~\ref{fig:Figure2}. However, shortcuts stops ``accelerating'' the first-passage process at a $N$-dependent threshold value $r_0^*$, beyond which the MFPT develops a \emph{nontrivial structure}, namely, it emerges a window of $r$ values for which the MFPT exhibits a complex landscape of local maxima and minima, see the bottom panel of FIG.~\ref{fig:Figure2} for $d=N/2$, and ring sizes $N=5\times10^2$, $10^3$, $2\times10^3$, $5\times10^3$, $1\times10^4$, $2\times10^4$, $5\times10^4$ and $10^5$. The minima for this case ($d/N=0.5$) corresponds to the cusps of the darken-shadowed lines given by Eq.~\eqref{eq:d_minimas} and shown in the top panel of FIG.~\ref{fig:Figure2}, located at $r/N=(\frac{1}{2}-\frac{1}{N})\frac{\mu}{\nu}$ with $\mu=1,3,\ldots\le(N/2-1)$ and $\nu=1,\ldots,\frac{N}{2}-1$. This nontrivial structure ends at another $N$-dependent characteristic threshold value of $r$, $r_f$, beyond which the MFPT resumes its monotonic decay with $r$. We report $r_f^*=N/2-r_{f}$ instead of $r_f$ since a similar quantity is employed in \ref{AppendixB} to study monotonically intervals.

From a spectral perspective, the two independent cosine contributions in Eq.~\eqref{eq:eigenvalues}, associated with local and long-range transitions, compete generating oscillatory structures  as a function of the shortcut length. Since the MFPT depends on the inverse spectral gap and on weighted sums of eigenvectors, these oscillations propagate into the first-passage statistics, producing the observed sequence of maxima and minima. In this sense, the MFPT landscape can be interpreted as a resonance phenomenon in the spectrum of the stochastic operator.

The nontrivial dependence of the MFPT as function of $r$ is manifested as a \emph{self similar pattern} (see bottom panel in Fig.~\ref{fig:FigureA7} in the Appendix~\ref{AppendixA}), conspicuously conveyed in the landscape of maxima and minima shown in the bottom panel of FIG.~\ref{fig:Figure2}. The patterns can be interpreted as a manifestation of the hierarchical structure of commensurability relations between the shortcut length $r$ and the system size $N$. As the ring size increases, new rational relations between these quantities appear, generating additional extrema in the MFPT. This hierarchical organization naturally leads to a nested structure of extrema, which is captured quantitatively through the box-counting analysis performed in Appendix~\ref{AppendixA}. The resulting fractal dimension, which lies between one and two, indicates that the MFPT landscape possesses a nontrivial geometric structure in parameter space. Such behavior suggests that transport processes in networks with long-range connections may exhibit multiscale properties that are not captured by conventional mean-field descriptions.

Although the explored range of $N$ does not allow us to determine conclusively the asymptotic limit, the persistence of the nested structure over several orders of magnitude strongly suggests that the self-similar behavior is not a finite-size artifact but rather a consequence of the hierarchical organization of commensurability relations between $r$ and $N$.

\begin{figure}
    \centering
    \includegraphics[width= \columnwidth]{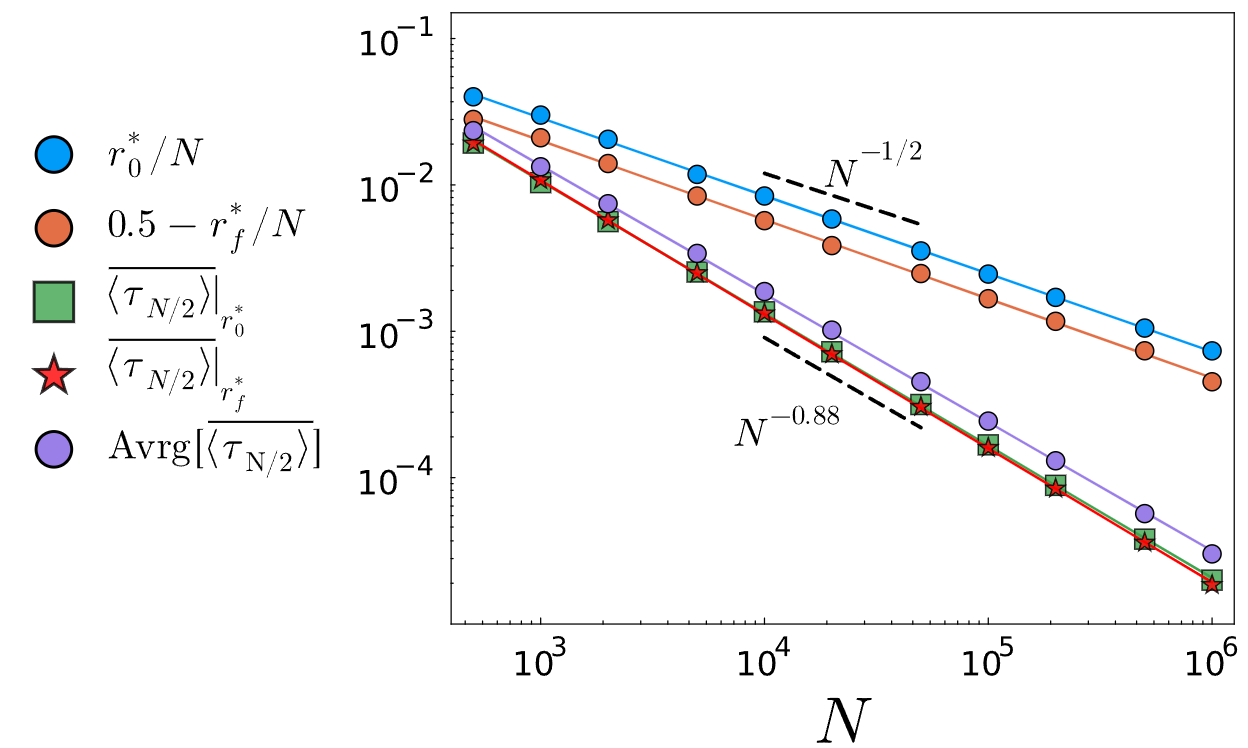}
    \includegraphics[width=\columnwidth]{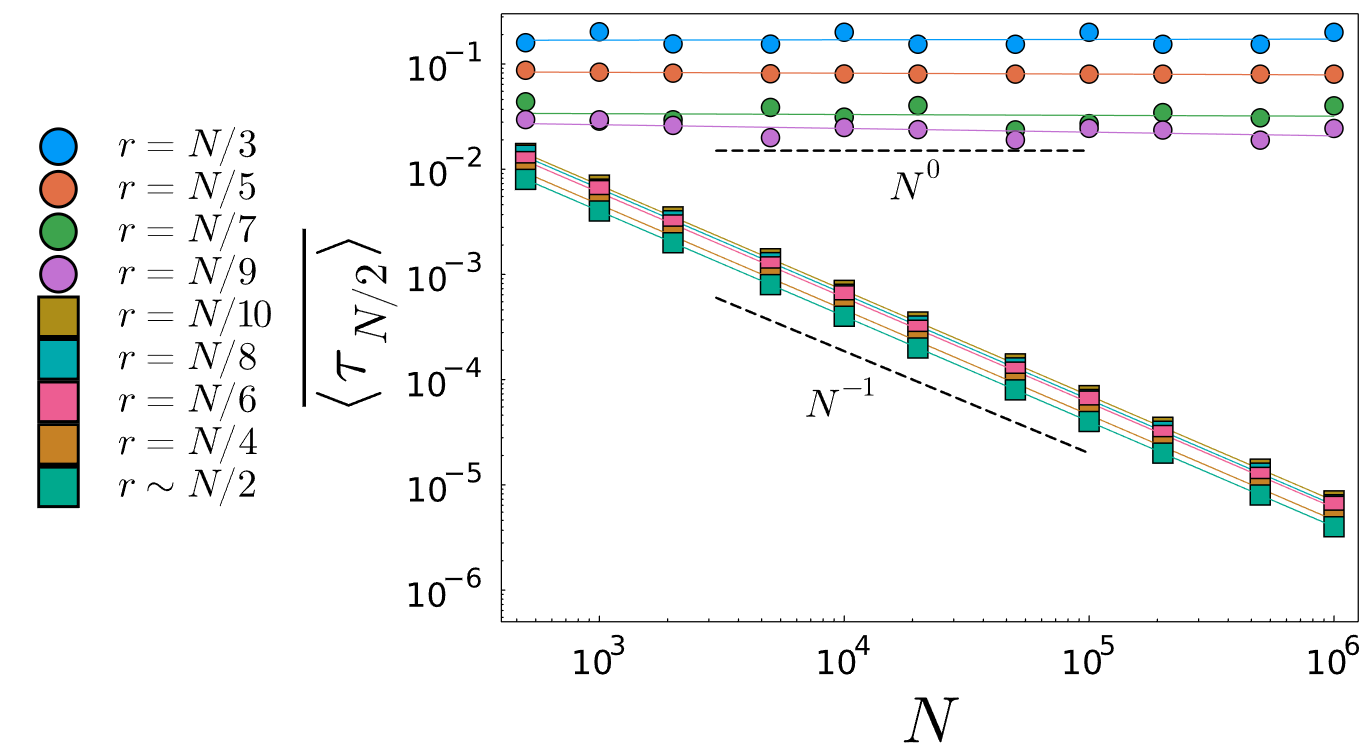}
    \caption{Top panel.- The scaling dependence of $r^{*}_{0}/N$, $r^{*}_{f}/N$, $\overline{\braket{\tau_{N/2}}}|_{r_{0}^{*}}$, $\overline{\braket{\tau_{N/2}}}|_{r_{f}^{*}}$ and Avrg$[\overline{\braket{\tau_{N/2}}}]$, on $N$. Bottom panel.- Scaling of $\overline{\langle \tau_{N/2}\rangle}$ with system size $N$ for different shortcut lengths $r$ (the label $r\sim N/2$ corresponds to $r=N/2-1$). Two distinct scaling relations are clearly identified whether $\overline{\braket{\tau_{N/2}}}$ is a maximum or a minimum. For the maxima we have $\overline{\langle \tau_{N/2}\rangle}$, color dots for $r=N/3$ (blue), $N/5$ (orange), $N/7$ (dark green) and $N/9$ (purple), barely depends on $N$, while for the minima the scaling is $\propto N^{-1}$, color squares for $r=N/10$ (brown), $N/8$ (aquamarina), $N/6$ (pink), $N/4$ (ocre) and $N/2-1$ (green).}
    \label{fig:Figure3}
\end{figure}

The dependence on the system size $N$ of the MFPT between a pair of nodes, is thus highly non-trivial. In the absence of shortcuts the MFPT scales as $N^2$, thus diverging quadratically as $N\rightarrow\infty$. This result is related to the divergence of the MFPT between two arbitrary locations visited by a one-dimensional Brownian particle in continuous time and infinite space \cite{Redner2001}.\\
When considering shortcuts, the scaling of the MFPT with $N$, for pairs of nodes separated by $N/2$, is analyzed in three different intervals of $r/N$ defined by the two characteristic values $r_0^*<r_f^*$. First notice that these boundaries scale with $N$ approximately as $N^{\alpha}$, with exponent $\alpha\approx-\frac{1}{2}$, as is deduced from the top panel of FIG.~\ref{fig:Figure3}. The MFPT evaluated at these boundaries, $\overline{\langle\tau_{N/2}\rangle}\vert_{r_0^*,r_f^*}$ scale with $N$ as $N^{\beta_{0}}$ and $N^{\beta_f}$ respectively,  with $\beta_0\approx-0.90$ and $\beta_f\approx-0.91$, as is shown in the top panel of FIG.~\ref{fig:Figure3}.

(I) \emph{For $r/N\in(0, r_0^*/N]$}  
the normalized MFPT $\overline{\braket{\tau_{N/2}}}$ decreases monotonically with $r/N$ as $r^{-2}$ (see top panel of FIG.~\ref{fig:FigureB8}), and therefore $\braket{\tau_{N/2}} \propto N^2/(2r)^2$ keeping the same $N^2$ scaling as in the absence of shortcuts (these intervals can be clearly noticed at the beginning in the bottom panel of FIG.~\ref{fig:Figure2}), however modulated by the value of $r$. This last result is in agreement with the relaxation scaling given by the relaxation time $\tau_\star$ pointed out in the previous section.

(II) \emph{For $r/N\in[r_f^*/N,0.5)$}, the MFPT decreases monotonically again, but with a different $N$-scaling relation. If $r^\prime=\frac{N}{2}-r$ measures the length of the shortcut from from the boundary $N/2$, the MFPT decreases as $\frac{N}{4}{r^{\prime}}^{0.35}$.  

(III) \emph{We focus now to the interval $r_0^*/N\le r/N\le r_f^*/N$,} where the MFPT exhibits a highly non-trivial structure. First notice that the average MFPT in this interval,
\begin{equation}
\label{MFPT-Avrg}
\text{Avrg}[\langle\tau_{N/2}\rangle]=\frac{1}{|R|}\sum_{r\in R}\overline{\langle\tau_{N/2}\rangle},
\end{equation}
approximately follows the same scaling as the MFPT at the boundaries, i.e. $\overline{\braket{\tau_{N/2}}}|_{r_{0}^{*}}\sim N^\beta$ , $\overline{\braket{\tau_{N/2}}}|_{r_{f}^{*}}\sim N^\beta$, and $\text{Avrg}[\langle\tau_{N/2}\rangle]\sim N^\beta$, this is shown in the top panel of FIG.~\ref{fig:Figure3} (green squares, red stars and purple dots, respectively) with $\beta\approx-0.88$. In Eq.~\eqref{MFPT-Avrg} $R$ is the set of shortcut lengths $r$ such that $r_0^*<r<r_f^*$, while $|R|$ its size or cardinality.\\
The minima and maxima of  $\braket{\tau_{N/2}}$ exhibit a distinct scaling with $N$. At the shortcut lengths $r$, where the MFTP exhibits maxima, the scaling relation with $N$ is quadratic with $\braket{\tau_{N/2}}\sim N^2$, as can be deduced from the fact that at the maxima $\overline{\braket{\tau_{N/2}}}|_{r_{0}^{*}}$ barely depends on the system size (see colored dots for $r=N/3$, $N/5$, $N/7$ and $N/9$ in the bottom panel of FIG.~\ref{fig:Figure3}, where the some of the most prominent maxima appear as shown FIG.~\ref{eq:Figure2}). In contrast, the MFPT minima scales linearly with $N$, i.e. $\braket{\tau_{N/2}}\sim N$ as can be deduced from the scaling shown in the bottom panel of FIG.~\ref{fig:Figure3}, where $\overline{\braket{\tau_{N/2}}}\sim N^{-1}$ (see color squares for $r=N/10$, $N/8$, $N/6$, $N/4$ and $N/2-1$).   \\

\subsection{Mean Square Displacement}

We now present an analysis of the effects of shortcuts on the mean squared displacement $\braket{x^{2}(t)}$ of the random walk considered here. The MSD is defined as
\begin{equation}
\label{MSD}
\braket{x^{2}(t)}=\sum_{x=1}^N (x-1)^2 P_{1,x}(t),
\end{equation}
where, without loss of generality, we have chosen the starting point of the walks at node 1. For rings of size $N$ the MSD converges asymptotically to
\begin{align}
    \braket{x^2}=&\sum_{x=1}^N (x-1)^2 P_x^\infty\nonumber\\
    =&\frac{(N+1)(N+2)}{12}.
\end{align}
In the large-size system regime, $N\gg1$ we have the characteristic value $\braket{x^{2}}_{sat}=N^{2}/12$.

After substitution of Eq.~\eqref{eq:probabilidad_s_j} into Eq.~\eqref{MSD} and some standard and straightforward calculations, we obtain in the short-time regime the standard linear time dependence of the MSD
\begin{equation}
    \braket{x^{2}(t)}\approx\frac{(1+r^{2})}{2}t,
    \label{eq:msd}
\end{equation}
where $r^2$-dependent diffusion coefficient $(1+r^{2})/4$ is identified. This takes into account the contribution from diffusing along links, $1/4$, and from diffusing along the shortcuts $r^2/4$. 

\begin{figure}
    \centering
    \includegraphics[width=\linewidth]{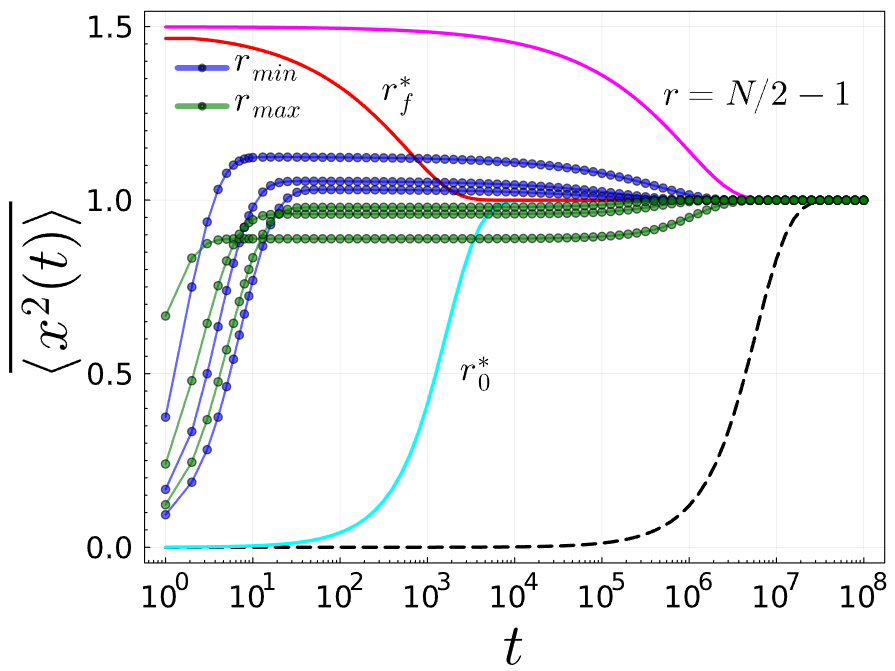}
    \caption{The MSD time dependence $\overline{\braket{x^{2}(t)}}$, for rings of size $N=10^{4}$ and representative shortcut lengths $r$. The case without shortcuts is marked by the dashed-black curve. Curves labeled with $r_{min}$ correspond to the values $N/4,~ N/6,~ N/8$ (values for which $\braket{\tau_{N/2}}$ attains minima) display a fast initial growth that overshoots the saturation level, followed by a slow relaxation back to it; the same overshoot relaxation pattern appears for $r=r_f^{*}$ (red line) and for $r= N/2-1$ (magenta line), because long shortcuts let the walker traverse the ring within a few steps. For $r_{\max}=\{N/3, ~N/5, ~N/7\}$ (values of $r$ for which $\braket{\tau_{N/2}}$ attains maxima) the MSD shows a long quasi-stationary plateau below the saturation level before converging to it. At $r=r_0^{*}$ (cyan line) the MSD grows monotonously.}
    \label{fig:Figure4}
\end{figure}

In Figure \ref{fig:Figure4}, $\overline{\braket{x^{2}(t)}}\equiv \braket{x^{2}(t)}/\braket{x^2}_{sat}$ is shown for some distinct values of $r$ as function of the steps $t$ for a size system $N=10^{4}$. The MSD exhibits the well-known monotonous growing behavior with $t$ for $2\le r\le r_0^*$ shown in Fig.~\ref{fig:Figure4}. In this range of values of $r$, the effect of shortcuts is to increase the growing   rate (as reference we have included the case for which no shortcuts are considered, see dashed-black line). 

In contrast, for values of $r$ in the self-similar region, $r_0^*<r<r_f^*$, the MSD exhibits two distinct growing behaviors depending on whether $r$ corresponds to local minimum ($r_{min}$), or a local maximum ($r_{max}$), of the MFPT. In both cases the shortcuts induce an ``fast'' growing behavior of the MSD which  evolves to a transiently saturating regime that depends on $r$ (see blue lines with dot symbols for $r_{min}=N/4=2500$, $N/6 \approx 1667$, $N/8=1250$, and green lines with dots symbols for $r_{max}=N/3\approx 3333$, $N/5=2000$, $N/7\approx1429$). Eventually the MSD relax to the asymptotic saturation value $\braket{x^2}_{sat}$. This transient saturation appears for particular shortcut lengths that allow for the recurrent traveling along trajectories of a characteristic length. For $r_{min}$'s, the transient saturating value exceeds $\braket{x^2}_{sat}$ while for $r_{max}$'s remains below it. 

Finally, for $r_f^*<r<N/2-1$ an ``explosive'' traveling is observed due to the long-range ``flight'' induced by the long shortcuts. The relaxation rate to $\braket{x^2}_{sat}$ depends on $r$ being larger for as $r\rightarrow N/2-1$ (see red line for $r_f^*$ and magenta line for $r=N/2-1$ in Fig.~\ref{fig:Figure4}). 

\section{The effects of stochastic resetting}
\label{sec:stochastic_resetting}
In the last years there has been a great interest of the effects of \emph{stochastic resetting} on the transport properties of random motion \cite{EvansPRL2011,EvansJPhysA2020}, particularly problems involving random walks \cite{MendezPhysRevE2016,Riascos_2020,GodrecheJStatMech2022,BarbiniJPhysA2024,MichelitschChaos2025}. In its simplest formulation stochastic resetting process refers to the sudden restarting of a process to a predetermined state at random times usually exponentially distributed \cite{EvansPRL2011}. The renewal dynamics under consideration induces stationary non-equilibrium distributions. For a freely diffusing Brownian particle subject to stochastic resetting, the stationary state is characterized by an exponentially decaying probability density centered at the resetting position \cite{EvansPRL2011}. Stochastic resetting also has a profound impact on first-passage properties, in this case, the MFPT, which diverges in the absence of resetting, becomes finite \cite{EvansPRL2011}. In the following lines we present the effects of resetting on the stationary distribution and on the MFPT properties.     

For a discrete-time random walk, the resetting process is implemented by choosing to perform the random walks dynamics with probability $(1-p_{rs})$ or to restart to the node $s$ with probability $p_{sr}$ independently of its current location. The master equation is now specified by   
the transition matrix that takes into account the resetting process given by
\begin{equation}
    \Pi^{sr} \;=\; (1-p_{sr})\,\Pi \;+\; p_{sr}\,S,
    \label{eq:transition_SR}
\end{equation}
where $\Pi$ is the transition matrix without resetting given by \eqref{TransitionMatrix}, and the $S$ specifies the nodes to which the walker is translocated upon resetting, when resetting to a single node is given by $\delta_{s,j}$.

The effects of stochastic resetting on the MFPT is investigated using the same formalism that yields Eq.~\eqref{eq:tau}, with the crucial modification that the analysis now involves the spectral properties of $\Pi^{sr}$, which differ from the particularly simple spectrum of $\Pi$ presented in Eqs.~\eqref{EigenSystem}. Notwithstanding this, $\Pi^{sr}$ is irreducible and a well defined stochastic matrix, therefore its spectrum leads to the calculation of the site occupation probability and the stationary distribution. The eigenvalues $\eta_\mu$, and the left and right eigenvectors, $\ket{\Psi_\mu}$ and $\bra{\overline{\Psi}_\mu}$ respectively, can be written in terms of those without resetting given by Eqs.~\eqref{EigenSystem} (see Ref.~\cite{Riascos_2020})
\begin{subequations}
\label{eq:EigenRes}
\begin{align}
\eta_{0}(p_{sr}) &= 1,\label{StationaryRes}\\
\eta_{\mu}(p_{sr}) &= (1-p_{sr})\,\lambda_{\mu},\quad \mu=1,2,\ldots N-1,
\end{align}
which are independent of the node of resetting $s$. The right eigenvector of $\Pi^{sr}$ corresponding to the eigenvalue \eqref{StationaryRes} is given by \cite{Riascos_2020} 
\begin{equation}
\ket{\Psi_{0}(p_{sr})}= \ket{\Phi_{0}}    
\end{equation}  
which does not depend on the node of resetting $s$ either, while the other right eigevectors are
\begin{equation}
\ket{\Psi_{\mu}(s,p_{sr})} = \ket{\Phi_{\mu}}
+ \frac{p_{sr}}{(1-p_{sr})\lambda_{\mu}-1}\,
\frac{\braket{s|\Phi_{\mu}}}{\braket{s|\Phi_{0}}}\,\ket{\Phi_{0}},
\end{equation}
for $\mu=1,2,\ldots,N-1$. The left eigenvectors are given by
\begin{equation}
\bra{\overline{\Psi}_{0}(s,p_{sr})}= \bra{\Phi_{0}}
+ \sum_{\mu=1}^{N-1} \frac{p_{sr}}{1-(1-p_{sr})\lambda_{\mu}}\,
\frac{\braket{s|\Phi_{\mu}}}{\braket{s|\Phi_{0}}}\,\bra{\Phi_{\mu}}
\end{equation}
for $\mu=0$ and
\begin{equation}
\bra{\Bar{\Psi}_{\mu}(s,p_{sr})} = \bra{\Bar{\Phi}_{\mu}}
\end{equation}
for $ 
\mu=1,2,\ldots,N-1$.
\end{subequations}

From the spectral representation (see Eq.~\eqref{eq:probabilidad_s_j}) the stationary distribution $P^{\infty}_j(s,p_{sr}) 
= \braket{s|\Psi_{0}(s,~p_{sr})}\,\braket{\overline{\Psi}_{0}(s,p_{sr})|j},
$ is obtained in the limit $t\rightarrow\infty$. Here we present an analysis when node of resetting $s=1$, which by the translational invariance of the ring configuration, no loss of generality is incurred, in this case the stationary distribution is
\begin{equation}
\label{StatDist-SR}
    P_{j}^{\infty}(s, p_{sr})
    =\frac{1}{N}\Biggl[1+ \sum_{\mu=1}^{N-1} \frac{p_{sr} \exp \left({-\textrm{i} \frac{2 \pi (j-1)}{N}\mu} \right)}{1-(1-p_{sr}) \lambda_{\mu} }\Biggr].
\end{equation}
In contrast with Eq.~\eqref{eq:pj_srs}, stochastic resetting makes $P_{j}^{\infty}$ highly nonuniform and strongly depends on the probability of resetting $p_{sr}$ and on the length of the shortcuts $r$ through $\lambda_\mu$. 
\begin{figure*}
    \centering
    \includegraphics[width=0.45\textwidth]{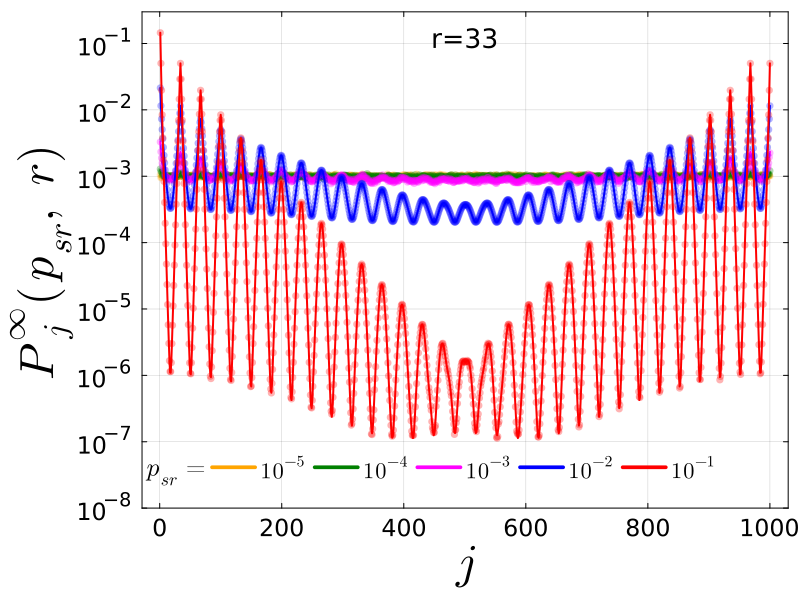}
    \includegraphics[width=0.45\textwidth]{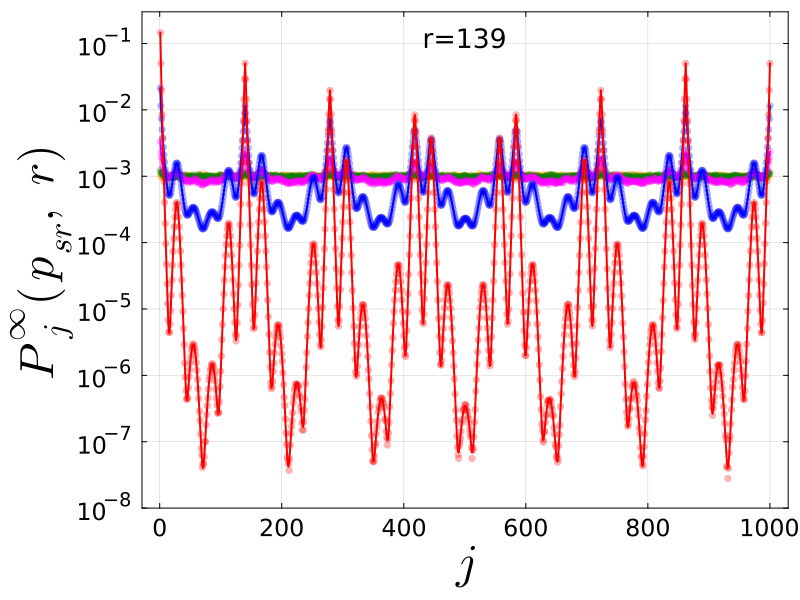}
    \includegraphics[width=0.45\textwidth]{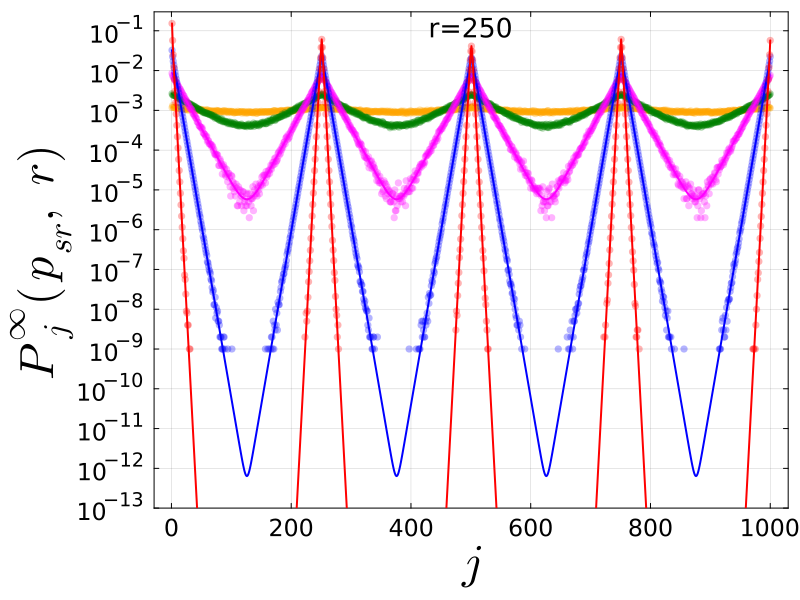}
    \includegraphics[width=0.45\textwidth]{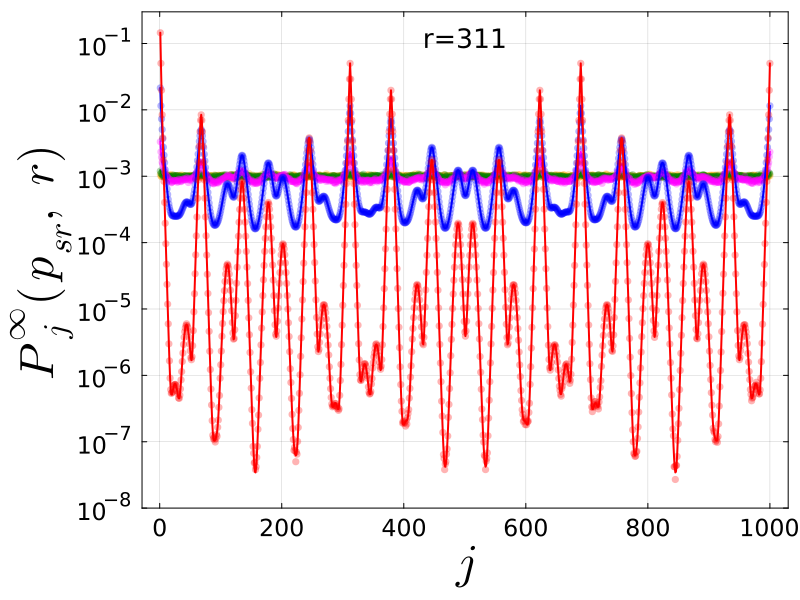}
    \caption{The stationary probability distribution $P_{j}^{\infty}(p_{sr},r)$, given in \eqref{StatDist-SR}, for sites $j$ on a $N=1000$ ring with shortcuts of length $r=33,~139,~250,~311$, and resetting probabilities $p_{sr}=10^{-5}$ (yellow line), $10^{-4}$ (green line), $10^{-3}$ (magenta line), $10^{-2}$ (blue line) and $10^{-1}$ (red line). Solid lines refer to the numerical evaluation of \eqref{StatDist-SR}, while  dots were obtained by Monte-Carlo simulations as explained in the text.}
    \label{fig:Figure5}
\end{figure*}
In Fig.~\ref{fig:Figure5} the stationary occupation probability of site $j$, $P^{\infty}_j$ given in Eq.~\eqref{StatDist-SR}, is shown for a ring of size $N=10^3$; probabilities of resetting $p_{sr}=10^{-5}$ (yellow line), $10^{-4}$ (green line), $10^{-3}$ (magenta line), $10^{-2}$ (blue line) and $10^{-1}$ (red line); and four different shortcut lengths $r=33$ (top-left panel), $r=139$ (top-right panel), $r=250$ (bottom-left panel) and $r=311$ (bottom-right panel). Solid lines mark the results obtained from Eq.~\eqref{StatDist-SR}, while dots mark the results obtained from an ensemble of $10^{6}$ final positions of trajectories of $10^{6}$ steps employing Monte Carlo numerical simulations ($p_{sr}=10^{-5},~10^{-4},~10^{-3}$); for the cases $p_{sr}=10^{-2},~10^{-1}$, numerical results were obtained after sampling $10^{9}$ positions obtained just before resetting events in order to improve computing times.  

It is clear that when resetting events are rare, $P_{j}^{\infty}(s, p_{sr})$ does not substantially differ from the uniform distribution \eqref{eq:pj_srs} (see the cases $p_{sr}=10^{-5}$, $10^{-4}$ and $10^{-3}$ in Fig.~\ref{fig:Figure5}). In contrast, for frequent resetting events, particularly for $p_{sr}\gtrsim 1/N$ (the smaller MFPTs in the system scale with the ring size as $\sim N$, particularly for the farthest nodes, thus the effects of resetting are conspicuous whenever $p_{sr}>N^{-1}$),
$P_{j}^{\infty}(s, p_{sr})$ becomes highly nonuniform as shown for the cases $p_{sr}=10^{-2}$ and $10^{-1}$ in Fig.~\ref{fig:Figure5}. For large enough values of $p_{sr}$, the sites neighboring the resetting site are conspicuously more visited, as a result, the stationary occupation distribution is peaked around the resetting site ($j=1$ in this case), while its first four neighbors: $j=2$, $N$, $1+r$ and $N-r$, are the second most visited sites, and so on. Red lines and red dots depict such a case for $p_{sr}=10^{-1}$. The values of $r$ and $p_{sr}$ are highly intertwined to produce the rich patterns shown in Fig.~\ref{fig:Figure5}. 

\begin{figure}
\includegraphics[width=0.93\columnwidth]{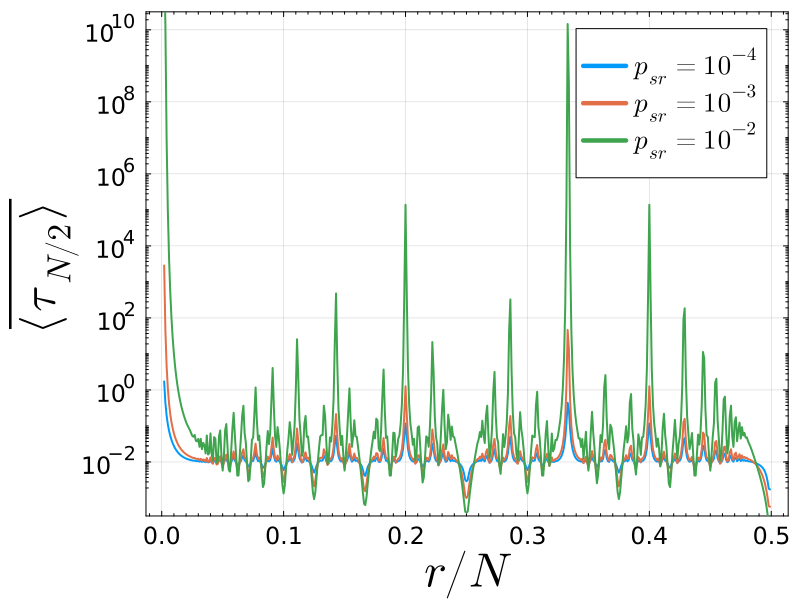}
\includegraphics[width=0.93\columnwidth]{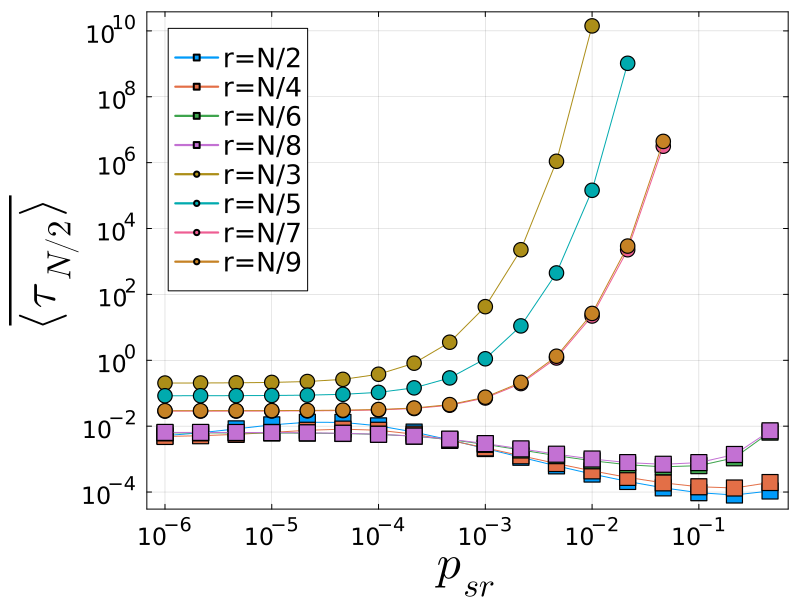}
    \caption{Top panel.- The MFPT $\overline{\braket{\tau_{N/2}}}$ between diametrically opposed nodes is shown for a network with $N=1000$ as function of $r/N$ and resetting probabilities $p_{sr}=10^{-4}$ (blue line), $10^{-3}$ (orange line) and $10^{-2}$ (green line). The oscillatory landscape is magnified as the resetting probability increases: maxima rise while minima deepen. Bottom panel.- The dependence of the MFPT $\overline{\braket{\tau_{N/2}}}$ on $p_{sr}$ is shown for values of $r$ that gives maxima ($N/9$, $N/7$, $N/5$ and $N/3$) and minima ($N/8$, $N/6$, $N/4$, $N/2$).}
    \label{fig:Figure6}
\end{figure}

It is well known how stochastic resetting affects the first-passage properties of brownian motion, making MFPTs finite when they diverge in the absence of resetting  \cite{EvansPRL2011,EvansJPhysA2020}. As Eq.~\eqref{eq:tau} establishes, the MFPT depends on the stationary distribution, $P^\infty_j$ which in turn is affected by stochastic resetting particularly in the frequent resetting regime. In Fig.~\ref{fig:Figure6} we show the normalized MFPT $\overline{\langle \tau_{N/2}\rangle}$ between the farthest nodes as a function of the shortcut length $r/N$ (top panel) for a ring of $10^3$ nodes and resetting probabilities $p_{sr}=10^{-4}$ (blue line), $10^{-3}$ (orange line) and $10^{-2}$ (green line). We observe that stochastic resetting exacerbates the nontrivial structure found in Sect.~\ref{sec:results_mfpt}, more precisely maxima becomes orders of magnitude larger and minima diminishes more with $p_{sr}$.  

In the bottom panel of Fig.~\ref{fig:Figure6}, $\overline{\langle \tau_{N/2}\rangle}$ is shown as function of $p_{sr}$ for particular values of $r$ is shown. 
The circles mark the cases for which the MFPT is a maximum at the particular values of  $r=N/3$, $N/5$, $N/7$ and $N/9$ in the figure. A fast grow of the MFTP with $p_{sr}$ is observed, i.e., the MFPT to reach the diametrically opposed node in the ring grows without limit as $p_{sr}\rightarrow1$. Resetting hinders the possibility of reaching that site. In contrast, for the values of $r$ for which $\overline{\langle \tau_{N/2}\rangle}$ is a minimum (see $N/2$, $N/4$, $N/6$ and $N/8$ marked in squares in the figure), the resetting probability induces a non-monotonous of the MFPT, exhibiting optimum values for $p_{sr}\sim 10^{-1}$. For numerical stability, only the values of $r$ whose corresponding $P_j^\infty$ exceed $10^{-15}$ are shown. 

\section{Conclusions}
\label{sec:conclusions}
We presented an analytical and numerical study of the mean first-passage time and mean square displacement for a random walk on a regular ring network of degree four with shortcuts of length $r$. Contrary to naive expectation, the presence of shortcuts does not always reduce the MFPT. As a representative result, the MFPT between the pair of farthest nodes exhibits a highly non-monotonic dependence on $r$, showing the emergence of a fractal-like self-similar structure within a specific range of $r$. In this regime, a sequence of maxima and minima in the MFPT clearly reveals the complexity induced by the network topology. 

Although the present study focuses on a deterministic ring network with shortcuts of fixed length, the mechanism responsible for the observed transport anomalies is more general since the phenomenon arises from the interplay between lattice periodicity and long-range transitions, a feature that appears in many systems displaying small-world characteristics. In this sense, the non-monotonic dependence of transport efficiency on shortcut length is expected to persist in broader classes of networks, including lattices with heterogeneous long-range connections or networks with multiple shortcut lengths. The ring topology considered here therefore provides a minimal model in which the fundamental mechanism can be isolated and understood analytically. 

From a practical perspective, the results obtained here reveal that the introduction of long-range connections does not necessarily guarantee faster transport across a network. Instead, certain shortcut lengths can dramatically hinder the first-passage process, producing MFPT values comparable to those of a purely local network. This observation highlights the importance of the spatial organization of long-range connections when designing networks intended to optimize navigation or communication. In particular, the results suggest that efficient transport may require avoiding shortcut configurations that generate resonant or commensurate structures within the network. To understand how the MFPT diverges with system size, we analyzed its scaling with the number of nodes $N$. We find that the MFPT scales linearly with 
$N$ at the minima (enhanced transport properties) and as $N^2$, characteristic of the cases without shortcuts, at the maxima, highlighting the strong influence of the shortcut length $r$ on transport dynamics.

Shortcuts also induce distinct behaviors in the time dependence of the mean square displacement, allowing for specific dynamical regimes to be selected through an appropriate choice of $r$. In particular, parameter values for which the MSD exhibits an initial explosive growth are of special interest. After this rapid transient, the MSD reveals a temporary trapping of the walk on an a ring of an smaller effective size when 
$r$ corresponds to values that produce maxima in the MFPT, or in larger effective rings when $r$ corresponds to minima in the MFPT.

The incorporation of stochastic resetting further reveals how non-equilibrium mechanisms can interact with network topology to shape transport properties. Resetting events introduce a renewal process that competes with the exploration dynamics of the random walk, effectively biasing the occupation probability toward the resetting node. Our results show that this mechanism amplifies the oscillatory MFPT structure induced by shortcuts, making maxima larger and minima smaller as the resetting probability increases. This interplay between resetting dynamics and network structure suggests that resetting protocols could be used as a control mechanism to tune transport efficiency in networked systems.

These results suggest possible directions for implementing shortcuts in different classes of complex networks in order to gain control on navigation properties,  as in communication networks, peer-to-peer systems and routing protocols where designing optimal long-range connections or avoiding resonant structures that increase latency is relevant. In transportation systems where highways act as shortcuts within local streets that form regular lattices, transport bottlenecks or
unexpected travel-time patterns could be observed, relevant for metro network design, traffic routing and logistics optimization. Our results may also connect to multiple fields where transport on networks is fundamental as in
reinforced random walks in networks \cite{MaJRSocInterface2012}, human mobility and urban transportation systems \cite{JiangPhysRevE2009,LinTransRevs2013,Xu-JiePhysRevE2026} and highly nonmarkovian random walks \cite{Guerrero-EstradaChaos2025}.

Remarkably, the eigenvalues dependence on cosine contributions involving two characteristic spatial frequencies $k/N$ and $kr/N$, suggests a connection with quasiperiodic lattice models. Certainly, when the ratio $r/N$ approaches irrational values in the large-system limit, the resulting spectral modulation becomes quasiperiodic. Similar mathematical structures arise in the Aubry-André model \cite{Dominguez-CastroEurJPhys2019},
where a quasiperiodic potential produces hierarchical spectral properties and localization phenomena. In the present stochastic setting, the interaction between these two frequencies leads to clusters of eigenvalues near unity, which strongly influence the mean first-passage time. The resulting hierarchy of extrema in the MFPT landscape may therefore be interpreted as a stochastic counterpart of quasiperiodic spectral interference.

\begin{acknowledgments}
This work was supported by UNAM-PAPIIT IN112623.
O.I.T.-M. acknowledges doctoral scholarship 676620 from SECIHTI (previously CONAHCYT).
\end{acknowledgments}

\appendix

\section{Self-similar region of the MFPT}
\label{AppendixA}

\begin{figure}
    \centering
    \includegraphics[width=0.8\columnwidth]{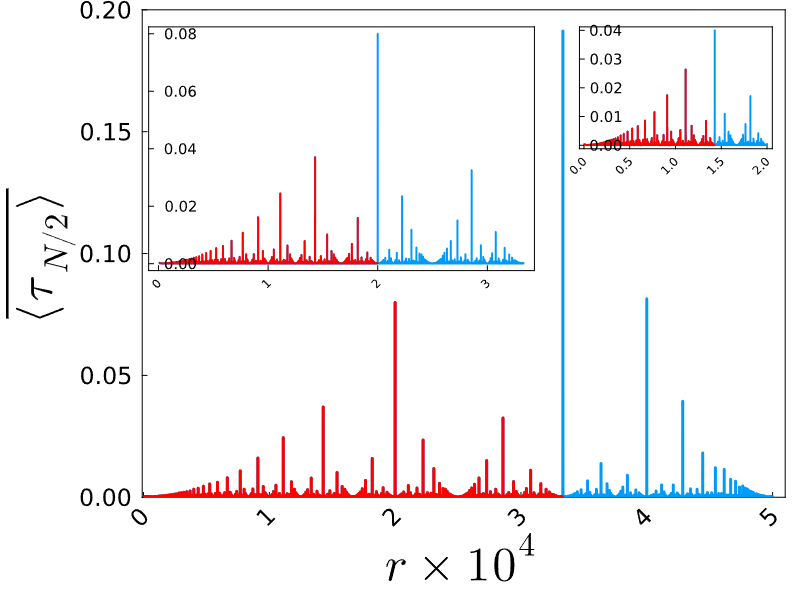}
    \includegraphics[width=0.8\columnwidth]{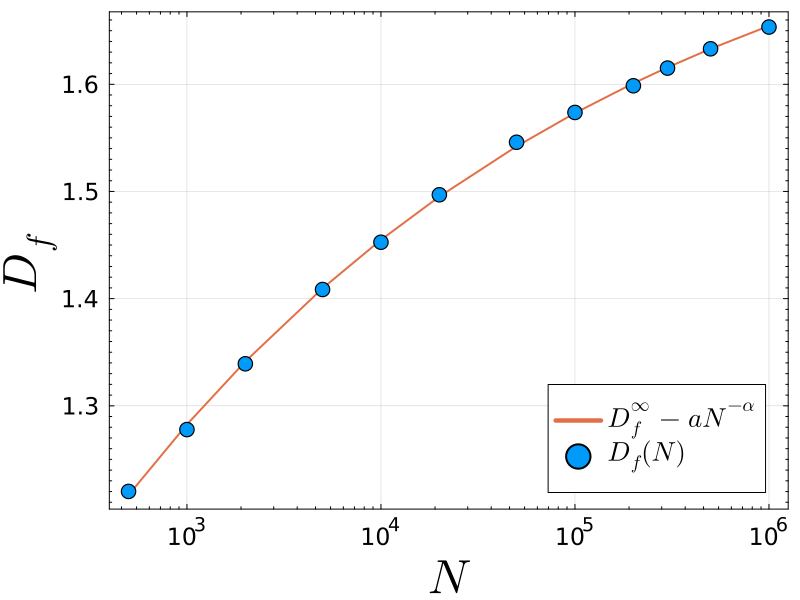}
    \caption{Top panel.- The self-similarity of the MFPT is shown for a ring size $N = 10^5$, where the main pattern is shown to approximately    repeats at the smaller intervals shown in the insets. The red/blue partitioning highlights a nested structure of subintervals associated with local extrema. 
    Bottom panel.- Box-counting fractal dimension $D_f(N)$ (markers) together with a finite-size scaling fit (orange line) of the form $D_f(N) = D_f^\infty - a N^{-\alpha}$, yielding an asymptotic fractal dimension $D_f^\infty = 1.833$ characterizing the fractal dimension of the pattern, with $a = 1.697$, and $\alpha = 0.163$.}
    \label{fig:FigureA7}
\end{figure}

To further examine the self-similarity suggested by the bottom panel of Fig.~\ref{fig:Figure2}, we analyze the antipodal MFPT $\langle \tau_{N/2} \rangle$ as a function of the shortcut length $r$ for various ring sizes. The top panel of Fig.~\ref{fig:FigureA7} displays $\overline{\langle \tau_{N/2} \rangle}$ on linear–linear axes, where the local maxima are clearly identifiable. Starting from the onset of the non-monotonic regime at $r = r_0^*$, we highlight in red the interval $[r_0^*, r_{m_1}]$ ending at the first maximum $r_{m_1}$, and in blue the complementary region. The interval $[r_0^*, r_{m_1}]$ is then enlarged in the first inset, and the same construction is iterated: we retain in red the subinterval $[r_0^*, r_{m_2}]$ ending at the next maximum $r_{m_2}$, and color the remaining part in blue. The second inset repeats the procedure once more. This nested partition provides direct visual evidence of a hierarchical, self-similar organization of the extrema.\\

A standard way to quantify self-similarity is the box counting  method. We apply box-counting method to the point set $\Omega_N$ formed by the sampled values of $\overline{\langle \tau_{N/2} \rangle}$ (at fixed $d = N/2$) within the non-monotonic window of the bottom panel of Fig.~\ref{fig:Figure2}. The resulting fractal dimension $D_f(N)$ is shown in the bottom panel of Fig.~\ref{fig:FigureA7}, where a clear dependence on $N$ is observed.\\

The finite-size dependence of $D_f$ is well captured by the leading correction to scaling form $ D_f(N) = D_f^\infty - a N^{-\alpha}$, which provides a bounded extrapolation to large $N$. For the data in Fig.~\ref{fig:FigureA7} we obtain $D_f^\infty = 1.833$, with correction parameters $\alpha = 0.163$ and $a = 1.697$. While other functional forms may also fit the available sizes, the power-law correction offers a parsimonious description of the observed slow convergence without introducing divergences at small $N$.

\section{Scaling up to $r_{0}^{\ast}$ and beyond $r_{f}^{\ast}$}
\label{AppendixB}
\begin{figure}
    \centering
    \includegraphics[width=0.8\linewidth]{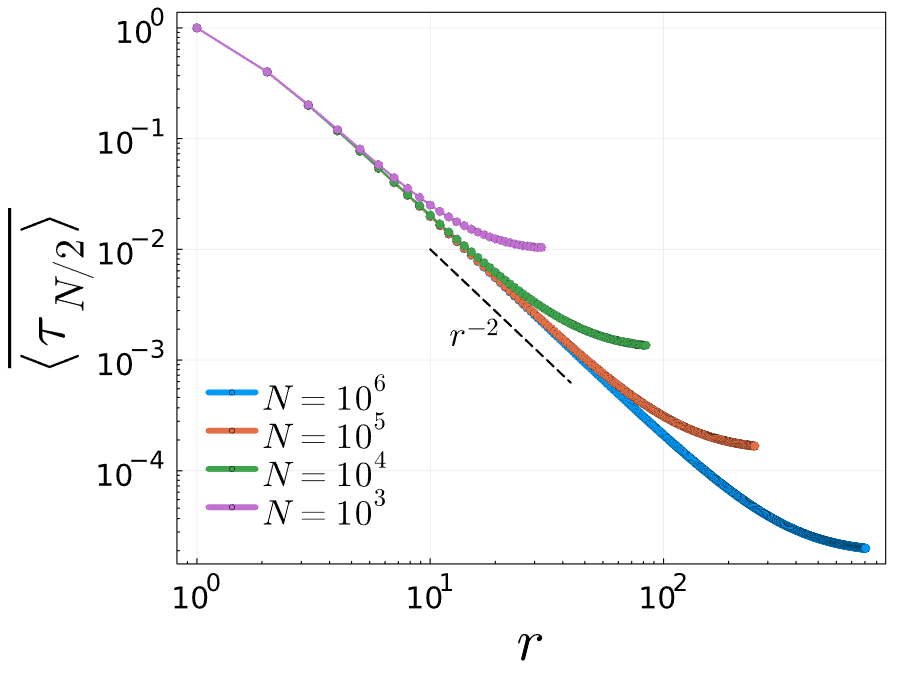}
    \includegraphics[width=0.8\linewidth]{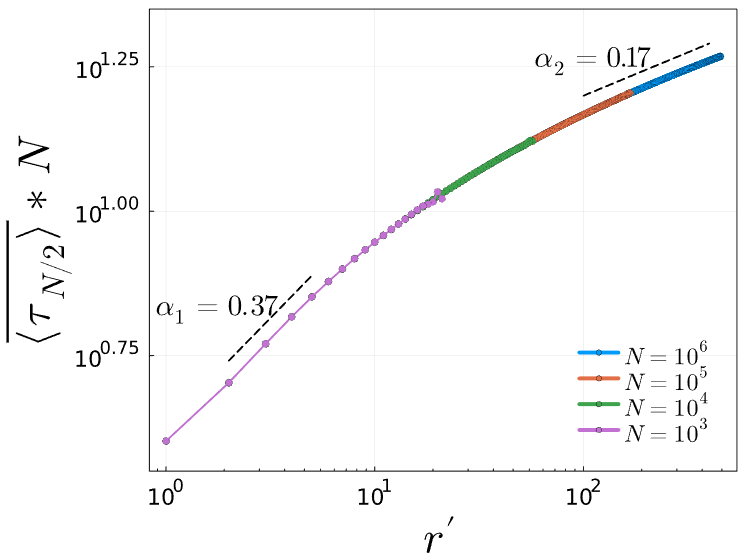}
    \caption{Monotonic branches of $\langle\tau_{N/2}\rangle$. Top panel.- small-shortcut regime ($r/N\ll1$); the dashed guide shows the $r^{-2}$ law. Bottom panel.- terminal branch near $r/N\to 1/2$. Plotting $\langle\tau_{N/2}\rangle,N$ against $r' = N/2 - r$ reveals two power-law ranges. Slopes $\alpha_{1}=0.37$ and $\alpha_{2}= 0.17$ are shown to comparison, the $r'$ dependence weakens as $r$ approaches $N/2$. Markers are perfectly overlapped, showing a unique and not approximately dependence.}
    \label{fig:FigureB8}
\end{figure}

The results shown here demonstrate that simply adding shortcuts does not guarantee a monotonic reduction of the MFPT. FIG.~\ref{fig:FigureB8} isolates the two monotonic branches that bound the oscillatory window seen in bottom of FIG.~\ref{fig:Figure2}, from $r=1$ up to $r_{0}^{\ast}$, and from $r_{f}^{\ast}$ towards the boundary $r/N\to 1/2$. In the small shortcut regime ($r/N\ll 1$), the data follows very closely a power law with slope $-2$, as indicated by the guideline.

For $d=N/2$ and $r/N \ll 1$, the scaling reads
\begin{subequations}
 \begin{equation}
\log\!\left(\overline{\braket{\tau_{N/2}}}\right) \propto 
\log\!\left(\frac{1}{r^{2}}\right), 
 \end{equation}
from which we obtain 
\begin{equation}
\braket{\tau_{N/2}} \propto \frac{N^{4}}{4\,r^{2}}.
\end{equation}
\end{subequations}
Thus, at the beginning of the curves the dependence on $r$ is strong: a one decade increase in $r$ reduces $\overline{\braket{\tau_{N/2}}}$ by approximately two decades.\\
In the bottom panel of FIG.~\ref{fig:FigureB8}, we plot $\overline{\braket{\tau_{N/2}}}\,N$ as a function of the distance to the boundary, $N/2-r$, for $r>r_{f}^{\ast}$. Let $r' \equiv N/2 - r$ denote this distance. Then, the terminal branch exhibits
\begin{subequations}
\begin{equation}
\log\!\big(\overline{\braket{\tau_{N/2}}}\,N\big) \propto \alpha\,\log r',
\end{equation}
leading to the scaling
\begin{equation}
\braket{\tau_{N/2}} \propto \frac{N}{4}\,r'^{\,\alpha},
\end{equation}
\end{subequations}
with an effective exponent $\alpha$ that depends on boundary $r^{\prime}\sim15$. Below that boundary $\alpha \sim 0.35$, in the other extreme $\alpha\sim  0.17$. In other words, the terminal monotonic behavior depends not only on $N$ but also on the proximity to the boundary through $r'=N/2-r$, being weaker for $r$ close to $N/2$.

\section{Different transition probabilities for links and shortcuts}
\label{AppendixC}
\begin{figure}
    \centering
    \includegraphics[width=0.8\linewidth]{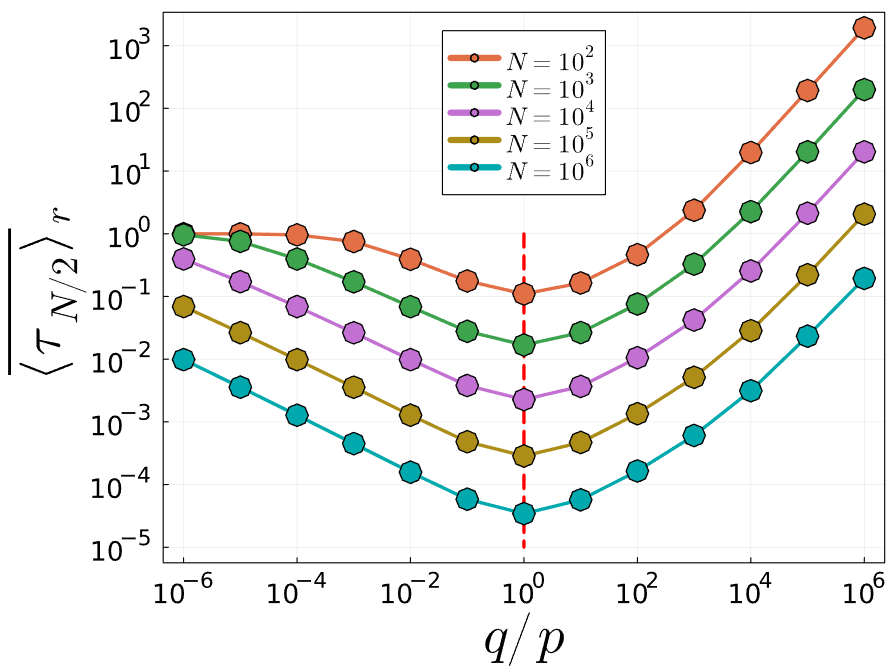}
    \caption{$\overline{\braket{\tau_{N/2}}}_r$ as a function on the bias ratio $q/p$ when links and shortcuts have different transition probabilities. The minimum occurs at $p=q$ (red dashed line). For $q/p<1/N^{2}$ the walker effectively ignores shortcuts, and the benefit of the shortcut architecture vanishes.}
    \label{fig:FigureC9}
\end{figure}

Throughout the main text we focused on the case where the walker has no preference between original links and shortcuts ($p=q$). More generally, one can bias the dynamics to prefer either shortcuts or links, which impacts $\braket{\tau_{s,j}}$. FIG.~\ref{fig:FigureC9} reports the average time on all $r$ ($\overline{\braket{\tau_{N/2}}}_r$), taken over shortcut lengths $r$, as a function of the bias ratio $q/p$, with $p$ the hop probability through original links and $q$ through shortcuts. The curve exhibits a clear minimum at $p=q$, indicating that, on average over $r$, the most convenient strategy for $d=N/2$ is to use links and shortcuts with equal likelihood.
While specific commensurate values of $r/N$ can drastically reduce $\braket{\tau_{N/2}}$ for a fixed shortcut length, averaging over $r$ reveals that the unbiased choice $p=q$ minimizes $\overline{\braket{\tau_{N/2}}}$.

\section*{Data Availability}

The raw and processed data that support the findings of this study are openly available in Zenodo at [10.5281/zenodo.19058589], reference number.

%


\end{document}